\documentclass[runningheads,a4paper]{llncs}

\usepackage{amssymb}
\usepackage{latexsym}
\usepackage{amsbsy}
\usepackage{amsmath}
\usepackage{natbib}
\usepackage{graphicx}
\bibpunct{(}{)}{,}{a}{}{;}
\usepackage{url}
\newcommand{\keywords}[1]{\par\addvspace\baselineskip
\noindent\keywordname\enspace\ignorespaces#1}

\begin{document}

\mainmatter  

\title{Big Data and the Internet of Things}

\titlerunning{Big Data and the IoT}

%
%
\author{Mohak Shah%
}
\authorrunning{Mohak Shah}

\institute{Research and Technology Center - North America\\
Robert Bosch LLC\\
Palo Alto, USA\\
\email{mohak@mohakshah.com}
}

%
%

\toctitle{Big Data and the Internet of Things}
\tocauthor{Mohak Shah}
\maketitle

\begin{abstract}
Advances in sensing and computing capabilities are making it possible to embed increasing computing power in small devices. This has enabled the sensing devices not just to passively capture data at very high resolution but also to take sophisticated actions in response. Combined with advances in communication, this is resulting in an ecosystem of highly interconnected devices referred to as the Internet of Things - IoT.  In conjunction, the advances in machine learning have allowed building models on this ever increasing amounts of data. Consequently, devices all the way from heavy assets such as aircraft engines to wearables such as health monitors can all now not only generate massive amounts of data but can draw back on aggregate analytics to ``improve'' their performance over time. Big data analytics has been identified as a key enabler for the IoT. In this chapter, we discuss various avenues of the IoT where big data analytics either is already making a significant impact or is on the cusp of doing so. We also discuss social implications and areas of concern.
\end{abstract}

\section{Introduction}

In recent years, technological advances have opened up entirely new opportunities for both collecting and processing large-scale data. The capability to build algorithms that can generalize and do inductive inference has also increased  significantly. This has resulted in advancing the state-of-the-art in traditional research fields that relied on huge quantities of data but were challenged by limited data acquisition capability or computing power. Research fields such as astronomy, physics, neurosciences, as well as medical genomics are some immediate examples (see, for example,~\citep{feigelson12,hesla12}). Further, largely driven by problems such as \emph{search} and then those pertaining to social media, novel data- and compute-architectures as well as learning algorithms have also appeared in recent years. This has further propelled the prospects of building value added offerings.

In conjunction, there have been immense developments in sensing technologies resulting in ``smart'' devices that are constituted of sensors, actuators as well as data processors. We are at the cusp of a revolution in terms of how humankind interact with the technology in that an ever-increasing number of devices that we use, operate or interact with (even passively) are capable of collecting these actions, and more, in the form of data. As~\citet{zaslavsky13} note, ``...concentration of computational resources enables sensing, capturing, collection and processing of real time data from billions of connected devices serving many different applications including environmental monitoring, industrial applications, business and human-centric pervasive applications.'' Such sensing technology is becoming pervasive and ubiquitous, and will be able to collect data through intermittent sensing, regular data collection as well as Sense-Compute-Actuate (SCA) loops. Hence, data can be collected at desired resolution all the way from continuous monitoring, to event or action captures. Moreover, such devices, be they appliances at home, heavy assets such as aircraft engines in the field, or wearables and mobile devices, do not function in isolation. More and more such devices or ``things'' are being ``interconnected'' resulting in an ecosystem referred to as the \emph{Internet of Things-IoT}. This interconnectivity offers opportunities for enhanced services and efficiency optimization that can supplement each other by means of derived and abstracted insights - higher level of observations and inferences made from data arriving from multiple interconnected devices. Note that this interconnectivity need not be a device-to-device or machine-to-machine interconnectivity but can also be achieved via common platforms. Moreover, this can both be (near-) real-time as well as passive (data collected and analyzed over time). 

Gartner estimates that, by 2020, this network of interconnected devices will grow to about 26 billion units with an incremental revenue generation in excess of \$300 billion, primarily in services. Furthermore, global economic value-add through sales into diverse end markets would reach \$1.9 trillion~\citep{gartnerreport13}. Consequently, the data resulting from these devices will grow exponentially too resulting in new business opportunities as well as posing novel challenges to managing and processing it for value gain. The data of the digital universe is slated to grow 10 folds by 2020. Various research and analysis firms have confirmed the scale of these projections in addition to the Gartner report. IDC further notes that data just from embedded systems, i.e. sensors and physical systems capturing data from physical universe, will constitute 10\% of the digital universe by 2020 (this currently stands at 2\%) and represent a higher percentage of target-rich data~\citep{emcidcreport14}. These technologies are also resulting in novel business models as well as new revenue sources, diversification of revenue streams in addition to increasing visibility and operational efficiency. Businesses will increasingly focus on services' aspects enabled by an increased understanding of utilization and operation of assets, consumer interests and behaviors, along with usage patterns and contextual awareness. Consequently, the IoT in specific contexts has also been referred to as the Internet of Things and Services (IoTS). It is also referred to as the Industrial Internet to highlight the applications in the world of heavy industrial assets. We will, however, stick to the general term IoT to look at the opportunities that cut across domains as well as services.

\subsection{Chapter focus}

In this chapter, we will review some important aspects of the intersection of big data analytics and the internet of things. Even though we will briefly discuss the connectivity, communication and data acquisition issues, this is not the main focus of the chapter. We would rather like to focus on the novel opportunities and challenges that the new world of interconnected devices offer, along with some advancements that are being made on various fronts to realize them. Importantly, we will also discuss social implications as well as some of the, possibly underappreciated, areas that need responsible consideration as we move forward with a technology with a profound impact on society. 

As a consequence of potentially billions of connected devices, the landscape of both handling and learning from data will undergo massive change. Further, the speed and scale at which the edge devices\footnote{Note that we use the term ``edge devices'' loosely to encompass not just the devices such as RFID tags but also other sensors esp. MEMS, including embedded sensors, monitoring and diagnostic sensors aboard industrial assets and so on.} will produce data will dwarf those of the current big data enablers such as social media, let alone manual data generation. This, previously unseen speed and scale of data, of course introduces challenges not only to the data and computing infrastructure but also pose a challenge to conventional learning methodologies and algorithms. As~\citet{aggarwal13} rightly note, \textit{scalability}, \textit{distributed computing} and \textit{real time analytics} will be critical for enabling the data-driven approaches to generate value. 

We would also like to reiterate the point made by~\citet{aggarwal13} that the concept of the internet of things goes beyond those of RFID technology and social sensing. While the former can be considered as a key enabler of the IoT, this technology is not the sole source of data acquistion as we noted above. Similarly, social sensing referring to peoples' interactions via embedded sensor devices, is a subset of IoT whereby this concept is not limited to people but also extends to machines and devices. Furthermore, we would also like to bring into discussion the resulting services based offerings that would be generated of this network. This is not just servitization, that refers to ``the strategic innovation of an organization's capabilities and processes to shift from selling products, to selling an integrated product and service offering that delivers value in use''~\citep{Vandermerwe89,Lee14}. In our view, IoT goes beyond integrated product and service offerings to enable novel business and revenue models. A variety of views on the IoT have been proposed based on different contexts.~\citet{aggarwal13} categorize these views in three broad categories: things-oriented vision (focusing on devices), internet oriented vision (focusing on communication and interconnectivity), and semantic oriented vision (focusing on data management and integration).\footnote{We do briefly cover some of these categories since they are indeed critical and the effectiveness of IoT applications and capabilities are highly contingent on effective solutions in these areas.}

We would like to discuss a \textit{functional vision of the IoT}, a vision where the resulting data and insights, and not the enablement mechanisms, plays a central role. From a functional perspective, we discuss the basic components of an enablement stack and also current and some future areas where we envision witnessing the immediate impact. It should be noted that it is impossible to cover a topic such as the IoT, even in the context of big data, in its entirety in a book chapter. The aim of this chapter is to familiarize the reader with how big data analytics is a major part of the IoT vision and will be \textit{a}, if not \textit{the}, key player in deriving business and societal value. Finally, big data does not refer only to volume aspect of data but also to the variety and velocity - the three important V's used to describe big data all of which pose novel challenges.
 
The rest of the chapter is organized as follows: We discuss major components of a big data analytics stack in the context of IoT in Section~\ref{sec:stack}. Section~\ref{sec:IoTareas} then details various domains that stand to benefit from big data and the IoT, followed by recommendations on what steps organizations need to take in order to harness this value in Section~\ref{sec:orgs}. We then focus on the social implication issues as well as areas of concerns in Section~\ref{sec:social} and present some concluding remarks in Section~\ref{sec:conclusion}.
 
\section{Big data analytics stack for the IoT}\label{sec:stack}

We highlight in this section the major areas relevant to enabling analytics to leverage the value from the IoT as well as allow a general model to scale. These are also crucial to the broader ecosystem that would allow for devices to house analytical capabilities themselves. Any IoT application, whether it manifests at the user level or cloud level would need existence of an end-to-end analytics stack to support its functionalities. The offerings from IoT applications will be contingent on how each of these building blocks are realized. Of course, the levels at which each of these components will play a role in any specific application is subjective but they are a necessary condition nonetheless.

\subsection{Data acquisition and Protocols}

Most of the data acquisition in the context of the IoT happens through edge devices. Edge devices are referred as such since these typically reside at the edges of the network. That is, they are present at the point at which either a human or an asset interacts with the rest of the network and it is through these devices that the initial data will be acquired and possibly re-transmitted back to the network. Examples include health sensors on patients, activity monitors such as Jawbone, control and advanced monitoring sensors on industrial assets, weather sensors, movement sensors in a home, visual, sonar and laser cameras aboard an autonomous vehicle, diagnostic sensors on appliances, sensors embedded on mobile devices, etc. Radio Frequency Identification (RFID) tags were one of the first mechanisms of acquiring such data but there have been other devices including sensors such as microelectro mechanical sensors (MEMS), mobiles and wearables that have vastly expanded the possibilities of large-scale, high-resolution data acquisition.

Efforts have been underway to establish proper channels to acquire and persist data collected from the edge devices. While currently there is no agreed upon protocol for such acquisition, domain-specific mechanisms are appearing. There is certainly a need for accepted protocols for communication for these devices both to each other and to a central capability such as cloud to enable aggregate analytics. Technologies involving wired or wireless communication of homogenous devices as well as capture and transmission of sensor data for storage and processing by applications are referred to, broadly, as Machine-to-Machine (M2M) technologies. Some companies are going the proprietary route while there have also been announcements of open-source efforts (e.g., Bosch, ABB, LG and Cisco's joint venture announced recently to cooperate on open standards for smart homes; see appendix). Similarly, there have been other joint efforts trying to bring more standardization to the IoT including Open Interconnect Consortium (OIC), AllSeen Alliance, Thread group, Industrial Internet Consortium (IIC) and IEEE P2413~\citep{Lawson14}. Developing an open-source ecosystem has its advantages since broader community can contribute to the efforts. Moreover, given that the user community is involved in the development, adoption becomes relatively easier and wide-ranging. Since these devices collect high dimensional and high frequency captures of device states, this will in turn also require high-bandwidth connectivity. For instance, an aircraft engine can send data through 10's to 100's of sensors at millisecond-resolution and can generate multiple GB's of data per flight.

\subsection{Data Integration and Management}

One of the most distinguishing aspects of the IoT is the fact that the data is acquired from a variety of sources. In order to provide useful services the data from edge devices typically needs to be combined with external data sources including business data, utilization data of assets, geographical data, weather data, etc. Consequently, the data quality and management issues also grow exponentially. Combining and analyzing heterogeneous data is a major challenge. Efforts have been made to standardize data characterization so that a communication protocol can be developed for data exchange. The Open Geospatial Consortium, for instance, has developed various such protocols under the Sensor Web Enablement initiative allowing for interoperability for sensor resource usage. Some of the standard interfaces proposed as a part of the initiative include O\& M (Observations and Measurements, to encode the real-time measurements from sensors), SML (Sensor Model Language, to describe sensor systems and processes), Transducer Model Language (TML, to describe transducers and supporting real-time streaming of data), Sensor Observation Service (SOS, standard web service interface for requesting, filtering and retrieving sensor system observations), Sensor Alert Service (SAS, for publishing and subscribing alerts from sensors), Sensor Planning Service (SPS, for requesting user-driven acquisitions and observations), and Web Notification Service (WNS, for delivery of messages or alerts from SAS to SPS). See~\citep{aggarwal13} for more details. Further, in order to integrate and annotate the sensor data, the World Wide Web Consortium (W3C) has initiated the Semantic Sensor Networks Incubator Group (SSN-XL) with a mandate to develop semantic sensor network ontologies. These efforts have constituted a big part of the semantic web effort, further defining ontological frameworks such as the Resource Description Framework (RDF) and the Web Ontology Language (OWL) that enable defining ontologies such as SSN (Semantic Sensor Network) and SWEET (Semantic Web of Earth and Environmental Terminology) to express identifiers and relationships in various contexts. 

Further, given that the data is acquired in real time and field settings, there are myriad of issues around missing values, skewness and noise. The high resolution temporal nature of such data further makes it difficult to align multiple sources as well as devise strategies to learn from them in conjunction with static data sources. In the context of assets, the data is also accompanied by derived attributes - ones whose values are calculated from the raw data using a conversion mechanism. However, the protocols for obtaining the derived quantities are not uniform or standardized even within a given domain let alone across domains. Data integration becomes more difficult since it requires the reconciliation of such derived quantities. In addition to formulaic data transformations, there can also be hurdles in data management arising from issues such as privacy and security resulting in deidentified and/or encrypted data. 

From a storage perspective, classical relational databases are no longer enough since the data is not only from disparate sources but it also appears, or needs to be organized, in native forms such as documents, graphs, time-series, etc. The whole paradigm around data organization that addresses the set of requirements around big data is broadly referred to as NOSQL (standing for Not Only SQL) databases. This includes columnar data stores such as BigTable, Cassandra, Hypertable, HBase (inspired by the BigTable); key-value and document databases such as MongoDB, Couchbase server, Dynamo and Cassandra (also supports documents); stream data stores such as Eventstore; graph based data-stores such as Neo4j and so on. Each of these have associated technologies for efficiently querying and processing data from respective stores and have unique advantages and capabilities. For instance, services such as Flume and Sqoop allow for ingestion and transfer of big data while languages such as Hive, Pig, JAQL and SPARQL enable efficient querying of big data in various forms including ontologies such as the RDF.   From an integration perspective, the classical approaches of business-to-business (B2B) data integration do not apply either since such data cannot generally be organized using a master database schema. A combination of these storage strategies are typically employed depending on the types of data and customized views can be created depending on the application requirements.

A data persistence strategy is also needed since many times storing such high resolution data in massive quantities is neither viable nor needed. Strategies involving data summarization and sampling along with storing accompanying metadata can be quite effective especially when the data has very high level of redundancies. Recall how sparse format allowed to store and process data files with few non-zero values much more efficiently in the case of very high dimensional data. These data structures, for instance, are a regular offering in various analytics toolsets and libraries such as pandas.

\subsection{Big data infrastructure}

The massive amounts of acquired data necessitates powerful infrastructure to support not just storing and querying, but also extracting insights from such data. Various categories of learning that need to be performed on such data exert unique set of requirements. For instance, one of the most common requirement is that of being able to perform batch analytics over historical data to build aggregate models. However, this can be a complicated endeavor given that the data does not necessarily reside on the same network let alone the same machine. Hence, parallel learning algorithms as well as distributed learning capabilities are needed depending upon the size, location and other data characteristics in addition to the communication constraints. Frameworks such as Hadoop have shown significant promise when it comes to distributed data analysis including efficient search, indexing as well as learning~\citep{aggarwal13}. Hadoop is a distributed storage and processing framework for large scale data relying on a Hadoop distributed file system (HDFS) with an aim to ``take compute to the data''. This is in contrast to the classical parallel high performance computing (HPC) architectures that relied on parallel file system where the computation would require high-speed communication mechanism to the data. Over past few years, Hadoop has developed as an ecosystem (see appendix) with various applications and services supporting functions on the core architecture allowing for efficient data storage and organization, search and retrieval (including querying), processing, as well as services such as resource scheduling and maintenance. Various data-stores as well as querying languages mentioned above form a part of this ecosystem.

One of the major limitations in the distributed settings such as Hadoop has been that of performing analytics with low latency requirements including model deployment, real-time, iterative, or interactive analytics. In such cases, especially when multiple passes on the data are required (e.g., many machine learning algorithms), Hadoop framework can be quite costly in terms of communication to the underlying HDFS. Frameworks such as Spark were developed to address these issues on Hadoop and since then have grown into its own ecosystem. These frameworks, especially Spark, have shown significant promise and are being investigated for their suitability in the IoT scenarios. Spark enables in-memory primitives for cluster computing as opposed to Hadoop's MapReduce which is a two-stage disk-based paradigm and hence allows for faster performance on applications with low-latency requirements mentioned above.  While both Hadoop and Spark offers streaming API's, Storm is a computational framework  designed with streaming analytics as its objective.  Since each of these paradigms have their strengths and limitations, choosing the right storage as well as computional paradigm involves an in-depth analysis of requirements for the use-case in which these would be employed. However, due to their open-source nature, there has been significant effort in promoting interoperability of these frameworks. For instance, both the Spark and Storm frameworks can operate on Hadoop clusters and hence provide for easy integrability. Hadoop commercial providers such as Cloudera and Hortonworks have also announced support for Spark and Storm respectively. Companies such as Databricks are already providing commercial version of the Spark framework.  Finally, to effectively deploy and scale the analytics models, standardization and benchmarking mechanishms for analytics are available that can allow for efficient communication of these models. Predictive Model Markup Language (PMML) provides one such mechanism. There have been successful commercialization of such standards from vendors such as Zementis that provides not just an encoding mechanism but a full deployment capability. This includes an execution and scoring engine, namely Adaptive Decision and Predictive Analytics (ADAPA) that can run PMML specified models allowing for modular and efficient model deployment. Capabilities such as Velox also target machine learning model management and serving at scale~\citep{Crankshaw14}. 

\subsection{Machine learning and Data mining}

The natural subsequence to the handling, management and integration of IoT data, is the actual insight discovery step which is the ultimate goal of the network. Even though each step starting from the data acquisition onwards poses a variety of challenges for the IoT, the ultimate value from these steps is realized only when useful and generalizable insights can be derived from this data. The current use of edge devices (at least in the consumer domain) seem to be predominantly point-use, that is, operationalization at the single user level. However, as more and more devices get interconnected this will inevitably change. In fact, there are various use cases where this is already visible as we will discuss in the next section. Learning from IoT data is particularly interesting and challanging at the same time. Classical machine learning methods need to be extended and adapted to cope with the challenge of scale, diversity and the distributed nature of the data. The volume, acquisition speed and temporal nature of the sensor and other related data is already highlighting the limitations of traditional approaches to learning. Some of the major challenges include learning in distributed settings, learning from very high dimensional, high resolution temporal data and learning from heterogeneous and complex data. Novel frameworks such as the alternating direction method of multipliers (ADMM)~\citep{boyd11} have appeared to enable optimization, a core functionality of many learning algorithms, in such distributed settings. Furthermore, advances have also enabled versions of successful machine learning algorithms such as topic modeling via Latent Dirichlet Allocation (LDA)~\citep{Wang09,Zhai12},  convolutional neural nets, Restricted Boltzmann Machines (RBM's)~\citep{Salakhutdinov-thesis,Dean12}, Support Vector Machines, Regression and so on (see, e.g.,~\citep{Mackey14,Pan14,Gonzalez14}) for large scale settings. Online versions of various classical learning algorithms have also appeared allowing for faster execution times on large datasets.

Consequently, this has also necessitated extensions of the evaluation approaches to the learning algorithms~\citep{japkowicz11} to be extended to large scale settings. Some promising approaches for resampling in large scale settings such as the bag-of-little-bootstraps (BLB)~\citep{Kleiner14} have appeared that also provides a theoretical framework characterizing them. In addition, there have been advancements in methods aimed at analyzing streaming data at scale for event prediction, change point detection, time-series forecasting and so on owing to the use cases that require online learning or where the models need to be adapted to evolving realities (see, for instance, ~\citep{lin03}). Feature discovery is also one of the issues that has resurfaced since it is no longer feasible for learning-features to be designed or discovered in conventional manner. Novel approaches are enabling automated feature discovery and learning in cases where generalized models can be built from extremely large distributed datasets. One of the most prominent developments has been in learning sophisticated networks and autoencoders via Deep Learning methods~\citep{Dean12}. Deep learning has shown significant promise in domains such as image classification, speech recognition and text mining~\citep{Krizhevsky12,Socher11,Le12,Bengio03aneural}.

In addition to these, there have also been efforts to scale up the deployment of large scale classifiers in hardware and embedded systems. Specific chip designs inspired by both the machine learning and cognitive computing fields have appeared to this end. Some prominent examples include IBM's SyNAPSE, NVidia's Tegra X1 and Qualcomm's Zeroth (see appendix for links).

\subsection{Bringing the building blocks together}

From an organization or application level, it is clear that an end-to-end IoT stack is needed. The components of this stack will include data acquisition right from the M2M layer, data-processing, data-sharing through interconnected network, insights' discovery and capability to relay results both to devices (for potential actions) as well as to (automatic or manual) decision makers. Various teams and companies are identifying the nature and structure of such an IoT stack that can provide infrastructure, platform and services for both front-end application and solution development, as well as back-end computing and support (e.g., via cloud). Commerical vendors such as EMC, Microsoft, Amazon and IBM offer building blocks of this stack that can be instantiated by organizations or service providers based on their specific requirements. Increased modularity and interoperability will further speed-up the adoption and scaling of these capabilities. For example, being able to choose the desired infrastructure, platform and software selectively from a combination of vendors can address specific needs of IoT applications. Consequently, Infrastructure-, Platform-, and Software-as-a-service (IaaS, PaaS and SaaS respectively) are becoming increasingly desirable (see, for example, offerings from Cloud Foundry, Microsoft Azure; link in appendix). Lambda architecture~\citep{marz15} has shown promise as a basis that allows for batch and real-time analytics together. This also allows to account for the volume, velocity and variety of big data. Lambda architecture already underlies many Hadoop and Spark instantiations. Architectures for specific cases, such as embedded systems and sensor networks, are also being proposed (see, for instance,~\citep{Gubbi13,Yashiro13,Tracey13,Sowe14}).

\section{Domains impacted by Big data analytics and the IoT}~\label{sec:IoTareas}

The applications within the IoT domains depend highly on the respective business drivers leading to multiple manifestations of business cases through such network. Various works have attempted to paint a picture of the application landscape for the IoT. For instance,~\citet{chui10},  categorize the applications in two broad categories: i) Information and Analysis, consisting of tracking behavior, enhanced situational awareness, and sensor-driven decision analytics; and ii) Automation and Control, consisting of process optimization, optimized resource consumption, and complex autonomous systems. Another categorization comes from~\citet{markkanen2014} who categorize these in five categories viz. predictive maintenance, product and service development, usage behavior tracking, operational analysis and contextual awareness. The~\citet{cognizantreport2014} also discusses some of the opportunities in the IoT. 

IoT potentially goes beyond the possibilities mentioned in above reports in that it will also enable sophisticated services capabilities as mentioned earlier. In this respect, another categorization is quite illustrative that divides these opportunities in consumer-facing and business-facing opportunities~\citep{Leuth14}. 

To provide a flavor of the type of some specific applications, let us look at some illustrative use cases from different IoT-related domains. Note that we are reviewing these domains from a big data and analytics perspectives. There are many more applications as a consequence of advancements in sensing technologies and hyperconnectivity achieved in the IoT. As the readers will notice, our categorization has overlaps with various above-mentioned efforts. However, looking at the applications and opportunities from a domain perspective can provide a more coherent picture.

\subsection{Manufacturing}

~\citet{Lee13} describes manufacturing as a 5M system consisting of Materials (properties and functions), Machines (precision and capabilities), Methods (efficiency and productivity), Measurements (sensing and improvement) and Modeling (prediction, optimization and prevention). In this context, additive manufacturing can be considered as a process for creating products using an integrated 5M approach. Recent advances have significantly improved sensing capabilities and data gathering around various aspects of this 5M system. However, in traditional-, as well as in many cases advanced-, manufacturing setup such information gathering had a preventive or control purpose and hence  didn't necessarily serve an analytics-oriented insight discovery objective. Even in traditional sense, it can be argued that big data has been utilized for quite some time especially in the context of modeling. However, this usage typically corresponds to modeling based on data under simulated or nominal conditions in which the product or manufactured industrial asset is run under a controlled environment. For a manufacturing process, big data can enable functions such as correlating controller and inspection data. When this is combined with the traditional overall equipment efficiency (OEE) providing the production efficiency status, insights into the relationship between performance and the cost involved in a sustained OEE level can be obtained. This is particularly timely as there is a significant initiative, referred to as \textit{Industry 4.0}, to increase digitization of manufacturing with a goal to build an \emph{intelligent factory}, with cyber-physical systems\footnote{Cyber-physical systems consist of computational and physical components that are able to perceive real-time changes as a result of seamless integration~\citep{NIST13}.} and the IoT as the basis. 

The opportunity landscape in the manufacturing domain is vast, in addition to the OEE and performance optimization. Big data analytics can help (and has started to do so) in areas such as cycle time reduction, scrap reduction, product defect detection (e.g., to improve quality ratios, or detecting products that may lead to quality issues later), identifying and resolving issues with machine failure and optimizing material and design choices (see~\citep{kurtz13,RBMongowhitepaper14} for some examples in various manufacturing domains). As smart factories move beyond sole control-centric optimization and intelligence, big data can enable further optimizations by taking into account interactions of surrounding systems as well as other impact factors. For instance, the production cycle quality assurance can benefit not only from the quality data of current cycle, but can also analyze quality data from the previous steps (e.g., quality and monitoring data from parts-suppliers) or feedback (e.g. quality reports and issue notifications from consumers). It should be noted that while such benefits result in immediate value, they also have significant indirect advantages. For instance, while slight increase in the quality as a result of improved defect detection may seem to be a marginal improvement for advanced manufacturing facilities, these can translate into new business opportunities for companies and sometimes can be a major deciding factor for the clients. Similarly, identifying defective or potentially defective parts right at the manufacturing or quality testing stages can mean reduction in quality claims at later stages. Such benefits have big multiplication factors in terms of business values associated with them, of course not to mention intangible benefits such as credibility and brand building for manufacturers. In addition to the above opportunities directly related to the manufacturing process, big data and predictive analytics can have a significant impact on making cyper-physical systems much more effective and efficient. Predictive analytics are also poised to address important issues in areas such as capacity planning due to uncertainty in downstream capacities, inventory and supply-chain management by reducing uncertainities around material and part availabilities, and by reacting to (or anticipating) market and customer demand changes. Importantly, this can also help understand and address product design and performance issues, and can help to complete the loop with respect to the manufacturing process at the material and design stages. This final aspect in fact has immense significance and leads us to the next area of operation and maintenance of (heavy industrial) assets.

\subsection{Asset and Fleet management}

Over the past few years, most organizations have undergone a major change in their business models or are in the process of doing so referred to as servitization. This basically emphasizes a customer focus in product and service delivery, and is already a major factor in consumer-oriented companies (e.g., home appliances and electronics). However, this has taken on even higher importance in asset-heavy organization such as manufacturers and operators of heavy assets like aircraft engines, locomotives, turbines, mining and construction equipments. The renevue models of these organizations have undergone drastic changes over past few decades. While manufacturers drew majority of their revenues from sale of these assets earlier, they do so now by selling service agreement and performance guarantees over the lifetime or usage of these assets. Moreover, such agreements rely heavly on asset utilization and hence it is imperative for the organizations that they have a very high visibility into asset operation so that they can not only quickly address but also effectively anticipate any major impending failure (at least at the asset level but ideally at the component level) that can jeopardize operational efficiency of the operators. Achieving an optimal efficiency and availability of asset are critical to both these manufacturers and their clients. Moreover, this also enables effective planning on the maintenance actions, performing fast and effective root-cause analysis as well as detecting and anticipating warranty issues as early as possible (this is analogous to the requirement at the manufacturing plants discussed above).

For instance, in the context of aviation, capabilities for predicting anomalies as well as prognostics can be potentially integrated in the flight controls. The IoT can further help in correlating these anomalies with additional information such as weather, particulate matter, altitudes as well as put this in context with fleet level statistics~\citep{Brasco13}. Big data acquisition capabilities are further enabling high resolution monitoring of aircraft engines. While traditionally field engineers were able to monitor snapshot data from engines, now full flight data can be reliably analyzed. Further, large scale analysis on multiple sensors can be performed that facilitate tasks such as event detection, signature discovery and root cause analysis. Other industrial domains stand to benefit from these approaches as well. Data is captured at different stages during the life and operation of assets. This data is organized in disparate forms and is typically disconnected over the stages and actions taken in maintaining and operating an asset. Some of these major data sources include condition monitoring data (present and historical), controller parameters, digitized machine performance data, machine and component configuration, model information, utilization and operational data, as well as maintenance activities. Leveraging such data would not only allow building of aggregate models for the fleet but also for more customized models for unique (set of) assets. Further, such information can be combined at the fleet level in order to understand aspects such as asset-deterioration patterns and behavior under varying operating conditions. Also, the underlying physical models that otherwise explain the behavior of assets under nominal conditions and effect of operations and utilization patterns on engine life, can be enhanced. In turn, this can significantly impact the maintenance of the engines taking us closer to condition based maintenance. 

Furthermore, data-driven insights would allow for enhanced capabilities to manage and contain unanticipated field events (e.g., asset failures or malfunctions) by enabling their localization, subsequent root cause analyses and identification of the most efficient resolution mechanisms as well as future design changes. Remote maintenance is an opportunity that is already being realized in some cases.~\citet{Lee14}, for instance, discusses a case study on the remote maintenance of Komatsu smart bulldozers used in mining and construction. Such capability would have significant impact on asset reliability, maintenance scheduling as well as reduction of unplanned downtimes~\citep{Lee14,Zaki14} with the ultimate goal of having self-aware and self-maintenance machines. This would in turn benefit fleet operations, scheduling as well as optimizations. 

Fleet-level analytics would also enable more efficient fleet management as well as better user experience, and hence is not limited to the heavy asset industry; connected cars and networked electromobility are examples. For instance, the BMW group, Bosch, Daimler, EnBW, RWE and Siemens have come together to take on the ``Hubject'' initiative (see appendix)  with the aim of optimizing electromobility through convenient access to a charging infrastructure. Connecting electromobility service providers, charging station operators, energy suppliers, fleet managers, and manufacturers, as well as utilizing analytics to provide value-add services (e.g., identify closest charging station, and suggest charging routines), is a demonstration of end-to-end IoT enabled capability for consumers. A~\citet{RBMongowhitepaper14} outlines another example use case in the context of mobility and automation using telematics data. Similarly, as cars evolve (they typically have the computing power of 20 PC's processing about 25GB's of data an hour~\citep{connectedcarmckinsey14}) to a connected world, they are moving beyond optimizing internal functions. The connected car initiative aims at developing the car's ability to connect to external network and not just enhance in-car experience but also to self-optimize its operation and maintenance.

Other advantages of connected vehicles will obviously be for fleet management and companies that rely on such vehicle fleet for their operations or even entire business models. Examples include postal and courier services, delivery industries, servicing companies (e.g. consumer appliance services) and so on that would look for connected vehicles to optimized different value drivers like gas consumption, route optimization, service time reduction, resource allocation and fleet efficiency.

\subsection{Operations management}

In the above subsection the focus of our discussion was mainly mobile assets. However, big data analytics capabilities are also impacting operations and maintenance of stationary assets such as energy turbines, and plants (production, generation and so on). Various monitoring, positional and control sensors in plants are enabling more effective plant maintenance, identifying sub-optimalities as well as safety and security. Predictive maintenance of machinery and plants can allow for higher availabilities, as well as better guarantees on quality and for reducing process variability~\citep{ibmwhitepaper11,kurtz13}. By intelligently instrumenting the plants with advanced sensors as well as integrating information from existing control and monitoring sensors, we can develop a better understanding of plant's operational status, anticipate and predict failures, and identify sensor correlations to understand (and in some cases ascertain) sensor interdependence as well as adjust working set-points for the equipments. This can enable improved stability in plant operations as well as a reduction in the high alarm rates that can hamper the operations.~\citet{garcia11} proposes such a monitoring model for the case of an oil processing plant. In addition to improvement in plant utilization (efficiency) and availability (less downtime), benefits will appear in terms of a reduction in operating costs as well as the ability to make real-time decisions.

Similarly commercial facilities maintenance (e.g., large buildings, campuses or company facilities) in broader context can also be seen as a part of this topic. However, most efforts in those areas are currently focused on energy optimization and hence we will cover those a little later.

\subsection{Resource exploration} 

Resource exploration industry such as oil and gas, mining, water and timber constantly face challenges in terms of finding renewable reserves of natural resources, and balance them with volatility in demand and price. The goals for these industries are often achieving a delicate balance of increasing production and optimizing costs while at the same time reduce the impact of environmental risks (e.g., reduction in carbon footprint). Among these various industries, the oil and gas industry, and particularly upstream sector of it is a complex business that rely heavily on data.\footnote{Oil and gas industry can be viewed in three different segments: Upstream (concerned with exploration, drilling/development and production), Midstream (concerend with trading, transportation and refining) and Downstream (concerned with bulk distribution and retail). Refining step has components in both midstream and downstream sectors. } Even before the advent of ``big data'', these industries have made use of data from various sources of information whether it is in the context of studying soil composition during mining, or monitoring deep sea or sub-surface assets' health through traditional prognostics and health management methods. However, these industries face new challenges as the data grows exponentially in volume, resolution (or speed of capture) and in variety. For instance, seismic data such as wide azimuth offshore data results in very high volumes. In addition to the seismic data, structured data comes from sources such as well-heads, drilling equipments and multiple types of sensors, such as flow, vibrations and pressure sensors, to monitor assets. This needs to be combined not only with drilling and production data but also those from unstructured data sources such as maps, acoustic data, image and video data and well logs. Further, these data are used, studied, analyzed and processed by various business segments that in turn generate a variety of derived data such as reports, interpretations and projections. The industry stands to gain deep and meaningful insights if these data can be efficiently and effectively \textit{managed}, \textit{integrated} and \textit{reconciled}. Like many other areas, the foremost challenges are those of data management, preprocessing and more importantly ascertaining data quality. Working successfully with such data sources can mean a significant increase in production, possibly at lower risks to the environment and safety, reduction in costs as well as speed to \emph{first resources}. 

While many companies in the oil and gas sector such as Chevron and Shell, have started looking into leveraging big data analytics, much more effort is needed to benefit from these opportunities in various areas.~\citet{Baaziz13} identify areas that stand to benefit specifically in the upstream oil and gas industry (see~\citep{feblowitz12,nicholson12,hems13,seshadri13} for further details):
\begin{enumerate}
\item \textit{Exploration:} Enhancing exploration efforts (e.g., by helping experts verify field analysis assumptions where new surveys are restricted by regulations); Improved operational efficiency by combining enterprise data with real-time production data; Efficient and cost-effective assessment of new prospects by more efficiently utilizing geospatial data; Early identification of potentially productive seismic trace signatures; and building new scientific models via insights discovery from multiple data sources (e.g., mud logging, seismic, testing and gamma ray).
\item \textit{Drilling and Completion:} Building more robust drilling models from current and historial well data and subsequent integration into the drilling process; Adaptive models to incorporate new data; Improved drill accuracy and safety by analyzing continuous incoming data for anomalies and event prediction; Reduction in Non Productive Time (NPT), one of the major concerns in the industry, by early identification of negative impacting factors of operations while increasing foot-per-day penetration; Models for optimal cost estimation; Predictive maintenance for increased asset availability, reduction in downtime as well as managed maintenance planning.
\item \textit{Production:} Mapping reservoir changes over time for adaptation of lifting methods for enhanced oil recovery (e.g. to guide fracking in shale gas plays); More accurate production forecasts across the wells for quicker remediation of ageing wells; real-time production optimization by allowing the producer to optimize resource allocation and prices; Increased safety through earlier anticipation and prediction of problems such as slugging and WAG gas breakthroughs;
\item \textit{Equipment Maintenance:} We covered this more broadly in the previous subsection. In the current context, this refers to preventing downtime, optimizing field scheduling as well as maintenance planning on shop floor.
\item \textit{Reservoir Engineering:} More accurate engineering studies and a better understanding of subsurfaces by more efficiently analyzing data and subsurface models.
\end{enumerate}

IoT related opportunities also exist in the midstream and downstream industries. For instance,~\citet{seshadri13} further identifies opportunities in environmental monitoring (e.g., by analyzing real time sensor data for regulatory as well as company control compliance; maintenance prediction based on pollution levels), reducing set up times at refineries by quicker crude assay analysis for oil quality prediction, and predictive and condition maintenance on assets in the transportation and refinement facilities (both in mid- and downstream). Similarly, opportunities also exist in retail optimization (e.g., gas station automation).

While the above opportunity areas are detailed in the context of oil and gas industries, these areas, opportunities and challenges are quite similar in other resource exploration industries too (see, for instance,~\citet{mindcommercellcreport2014} for a broader discussion).

\subsection{Energy} 

Energy is another sector that can be transformed as a result of big data analytics. While utilities have been identified as one of the biggest stakeholders, various other industries whether they are direct energy producers, services companies performing campus and facilities' energy management or sectors relying on and impacted by energy consumption can foresee significant potential in increasing production, improving energy demand prediction, reducing uncertainities in energy supply, better resource management through efficiency gains and energy waste reduction. These can yield benefits not just from a financial and efficiency perspective but can also be instrumental from an environmental perspective. For instance, a study on energy efficiency from McKinsey and Co. concluded that a holistic program could result in energy savings in access of \$ 1.2 trillion and a reduction in end-use consumption by 9.1 quadrillion BTUs while eliminating up to 1.1 gigaton of greenhouse gas every year by 2020~\citep{MITBizreport15}.

The opportunity areas in the energy domain that stand to benefit from big data analytics include but are certainly not limited to:
\begin{enumerate}
\item \textit{Asset and workforce management at utilities:} Increasing availability (e.g., in weather conditions like storms) by reduction in downtime and maintenance optimization as well as identifying potential hazardous situations, outage management, wind-farm management by turbine optimization both at an asset- and an aggregate-level, reducing energy thefts, etc.
\item \textit{Grid operations:} E.g., load forecasting and load balancing primarily for peak-shaving which is an immediate priority, outage management and voltage optimization, optimizing network and energy trading and incorporation of distributed smart grid components into the storage system, proactive management of distribution network, combining energy facilities into virtual power plants,\footnote{https://www.bosch-si.com/solutions/energy/virtual-power-plant/virtual-power-plant.html} incorporating renewable energy sources into the grid, increasing grid flexibility and scalability, optimization of energy production and supply as well as efficiency gains for utilities.
\item \textit{Transportation:} E.g., reducing energy consumption by dynamic pricing for road use and parking, frequent updated traffic information and route optimization.
\item \textit{Infrastructure:} Smart cities (smart parking, traffic monitoring and control, structural health systems); Infrastructure management, smart grids, etc. Please refer to~\citep{zanella14,bettencourt14,Byrnes14} for a broader discussion on the implications and uses of big data and IoT in cities.
\item \textit{Residential and Commercial facilities:} reducing energy consumption by optimizing utilization (e.g., of HVAC and heating operations), smart meters to track consumption and subsequent load balancing, smart appliances and lighting for need-based operation, reduction in energy waste by identifying energy holes and sinks; smart systems for water, lighting, fire, power, cooling, security and notifications resulting in cost savings, preventative maintenance of critical systems and in environmental benefits~\citep{Reddy14}.
\item \textit{Operations:} efficiency gains and cost reductions in company operations, e.g., identification of energy sinks such as running unutilized resources, optimization of energy usage in data centers, etc.
\end{enumerate}

Please refer to~\citep{orts14} for a discussion on many of these perspectives. Further,~\citet{MITBizreport15} covers a variety of opportunities as well as challenges as cities both grow and get ``smarter'' through more sensors, and advanced network and communication capabilities.

\subsection{Healthcare}

\citet{Reddy14} points out that one of the major changes in the healthcare domain as a result of IoT is the ability to monitor staff and patients, and the ability to locate and identify the status of healthcare equipment/asset resulting in improved employee productivity, resource usage and efficiency gains, and cost savings. Further, as~\citet{nambiar13} note, ``big data analytics can improve operational efficiencies, help predict and plan responses to disease epidemics, improve the quality of monitoring of clinical trials, and optimize healthcare spending at all levels from patients to hospital systems to governments''. Many edge devices have been introduced to patients in particular, and a wider population in general. For patients, these range from temperature monitors, blood glucose-levels monitors, fetal monitors, electrocardiograms (ECG), and even electroencephalography (EEG)\footnote{See \url{http://www.mybraintech.com/} .} devices. Not only this, efforts are underway to move beyond monitoring towards comprehensive health management. See, for instance, the Health Buddy device to capture daily activities and status of high risk patients.\footnote{\url{http://www.bosch-healthcare.com/en/us/products/health_programs/health_programs.html}.} In addition to the patient, an increasing section of healthy population are routinely using health and activity monitoring devices such as the jawbone and fitbit, as well as various applications through sensors on the mobile devices. Consequently, this also allows for various value-added application that can guide healthy living practices. Moreover, the fact that these devices are connected to the cloud also enables anonymized aggregate analyses at population segment levels as well as across other dimensions such as geographies and demographics. While this means a more healthy lifestyle for healthy populations, the ability to monitor patients' conditions on a continuous basis through these devices has very significant advantages. For instance, combining remote monitoring capability and distant communication technologies (e.g., low-cost video-conferencing), can allow for efficient remote healthcare in areas where direct access to medical personnel is difficult or time-consuming. Many other benefits such as reducing the number of at-risk patients, a reduction in readmission risk, epidemic monitoring, mobile healthcare for home and small clinics~\citep{Ghose12}, ambient assisted living~\citep{dohr10} and chronic patient monitoring~\citep{paez14} can be realized. There are also advantages in pharmaceutical drug trials, treatment effectiveness, as well as chronic disease management. Big data and the IoT can further allow for valuable insight discovery and knowledge extraction from personal health record, or PHR (current and historical) data~\citep{poulymenopoulou13}.

In addition to being able to provide higher quality of healthcare, providers will also benefit from the other previously discussed benefits of IoT including predictive maintenance and real-time asset monitoring capabilities resulting in high availability levels, reduction in operating costs as well as increasing supply chain efficiencies. See~\citep{Bui11,doukas12,zaslavsky13} for further related discussions.

One of the major concerns in such large scale data analyses is with respect to the data privacy and security, mainly the protected health information (PHI). Mechanisms to ensure proper data de-identification and anonymization as well as ascertaining data security in communication channels is an extremely high priority. While this also applies to areas discussed above, these issues are critical in the healthcare domain and if ignored can have serious implications for both individuals and population at large.

\subsection{Retail and Logistics}

We are moving towards a shopping experience in connected supermarket. One of the widely discussed IoT use cases involves the pre-specification of shopping lists that can be communicated to the superstore so that the checkout wait-times (one of the major problems in retail stores today) can be reduced. However, there are more interesting and compelling use cases that will be built on top of the resulting data that is generated. For instance, shoppers' purchase patterns can be mined for recommendation of relevant items, and in addition sales and discounts can be highlighted. Combined with social data and other preferences that a consumer may make available, a consumer-centric experience can be created tailored to consumers' unique preferences. These capabilities can also enable effective monitoring of shopper traffic across stores, targeted marketing as well as product positioning.

Further, being able to track product movement, e.g., through technologies such as RFID tags, will allow retailers to have a more accurate and efficient inventory management, increase inventory accuracy and reduce thefts as well as administrative costs. As~\citet{Reddy14} points out, big data capabilities along with IoT will result in stock-out prevention as a result of connected and intelligent supply chains, as well as real-time tracking of parts and raw materials allowing to preempt problems and address demand fluctuations. Naturally, this information can be fed back into the manufacturing and distribution channels for further optimizations leading to a reduction in required working capital, efficiency gains as well as avoiding disruptions.~\citet{RBMongowhitepaper14} illustrates another use case in retail where inventory can be tracked as it moves from shelf to basket allowing the retailer to enable analytics for optimizing available supply to predicted demand, reducing uncertainties and fluctuations through warehouse operations and the supply chain.

~\citet{waller13} discusses opportunities for big data analytics in general in the logistics industry highlighting potential areas of opportunity as real-time capacity availability, time of delivery forecasting, optimal routing and reduction in driver turnover, all when it comes to carrier optimization. More specifically, in the context of fleet management, logistics companies will rely increasingly on the capabilities offered by big data analytics and the IoT to harness benefits in various areas.~\citet{Jeske13} discuss areas relevant to the intersection of big data and IoT:
\begin{enumerate}
\item Optimization of service properties like delivery time, resource utilization, and geographical coverage - an inherent challenge of logistics.
\item Advanced predictive techniques and real-time processing  to provide a new quality in capacity forecast and resource control.
\item Seamless integration into production and distribution processes for early identification of supply chain risks leading to resilience against disruptions.
\item Turning the transport and delivery network, as a result of efficient sensor instrumentation, into a high-resolution data source. In addition to fleet management by network optimization, this data may provide valuable insight on the global flow of goods allowing the level of observations to a microeconomic viewpoint.
\item Real- or near real-time insights into (changes in) demographic, environmental, and traffic statistics by analyzing the huge stream of data originating from a large delivery fleet. 
\end{enumerate}

As IoT enables self driving vehicles, the logistics industry anticipates a large impact on end-to-end logistics operations as highlighted by~\citet{dhl-rb-kit-14}. Please also refer to \textit{Delivering Tomorrow}, for some studies on how such new trends are anticipated to impact the logistics industry in the future.\footnote{\url{http://www.delivering-tomorrow.com/}}

\subsection{Other opportunities}

In addition to the major sectors that we discussed above, IoT has many opportunities directly targeted to consumers. While the above discussion also identifies ultimate opportunities for consumers, such as higher quality and possibly more accessible healthcare for patients, better retail experience, smart homes and energy efficiency gains, the IoT will become a part of daily lives via direct interaction with a lot of devices; \textit{Wearables and Assistant devices} is one such important area. For instance, activity trackers such as jawbone and fitbit are already becoming a routine part of our lifestyle.~\textit{beddit}, \textit{withings Aura} as well as advanced versions of \textit{jawbone} can now also perform sleep monitoring, \textit{Being} does better tracking than regular accelerometers while devices such as \textit{Vessyl} can monitor what we drink. Connected wearables such as \textit{Ego LS}, a wearable camera, can stream live video while \textit{Tzoa} can do real-time environment tracking including pollution and UV exposure. Monitoring children is also made easier by devices such as \textit{Pacif-i} and \textit{Sproutling}. In addition, there are many assistant devices that intend to make our lives easier including home assistant robots (e.g., \textit{Jibo}), automatic lawn mowers (e.g., Bosch's \textit{Indego}), home interaction devices (e.g., Amazon's \textit{Echo}) as well as self-driving cars. These are of course just a few illustrative examples. There are a myriad of devices in the market today.

\subsubsection{Integrated Systems and Services}\label{subsec:integratedsys}

~\citet{Metz15} describes the wearables revolution of sorts as a result of IoT-enabled devices like the ones mentioned above. However, when it comes to consumers, we believe that most of the benefits from these developments will come not from individual offerings but from integrated systems. This requires not just connected, but interconnected devices. That is, the interaction doesn't just happen via the cloud but also between devices allowing them to adapt their bevahior as per requirements. For instance, a complete home automation systems that can not only control temperature, lighting and energy consumption of home appliances, but can also connect, communicate and coordinate with assistant devices such as \textit{echo}, vacuum or lawn mower as well as other aspects such as cars for a seamless and integrated experience to the user and optimization of resource use. One can similarly think of an integrated health management capability. Even though self driving or autonomous cars can be considered an exception here since they can be self-contained, they would also draw benefits from these capabilities. These would be beneficial in a range of areas includng transportation, logistics and mobility. Similarly, in industrial facing applications, this would mean more responsive, self-monitoring and potentially self-maintaining assets. For instance, wind-turbines can adapt their performance not just in relation to the wind and local weather but also in relation to the a global optimization at the aggregate level of a wind farm. Similarly, assets such as aircraft engines can be responsive in relation to their peers (e.g., assets operating under similar operational and utilization conditions). 

On protocols for data transfer and communication too, a variety of standards currently exist either due to various disparate efforts (to avoid dependencies) or due to companies developing proprietary offerings. Services companies will probably fill the gap created by non-existence of a common communication standard for various devices. The market will see a growth not only in such interconnection and integration services but also value-added services resulting from such integration. For instance, \textit{Tado} is providing an interfacing through a variety of heating systems from multiple manufacturers for a smart thermostat system. Beyond increasing interoperability of standards or devices, such services will also generate new business and revenue models and value-add capabilities allowing for better operations (e.g., improving availability, ensuring higher quality of service of systems~\citep{Deb13}), and financial risk modeling (e.g., better pricing and term-structure based on field operations).

Finally, many other sectors such as insurance (more informed risk modeling by utilizing real-time information), sustainability, social good, and security stand to gain with advancements in the big data and IoT technologies. The future holds even more promises such as opportunities with nano robots that can cure diseases, or in near-term, drones for various applications including deliveries and integrated surveillance functions. 

\section{Harnessing value: what do organizations need?}\label{sec:orgs}

From organizations' perspective, harnessing value not just from IoT related big data analytics, but data science in general, requires foundational capabilities to be set in place before useful insights discovery can begin. The \textit{analytics readiness} requirements include some of the capabilities discussed in Section~\ref{sec:stack} such as efficient storage and compute infrastructure, data acquisition and management mechanism, machine learning and data modeling capabilities as well as efficient deployment and scaling mechanisms. In addition, organizations also need to facilitate interfacing between engineering or domain experts and data scientists for efficient and productive knowledge transfer, agreed-upon validation as well as adoption and integration mechanism for analytics. 

There are three most important objectives that an organization needs to achieve to realize these gains as they transform to be more data-driven.
\begin{enumerate}
\item \emph{Data and Analytics strategies} that align with the business vision. While a lot of data science activities and modeling exercises can be done in a bottom-up fashion, a coherent strategy can guide how the individual scattered efforts come together. Such a strategy should take a comprehensive view of how analytics can be a part of the decision making and insights generation process in the light of existing as well as future business directions, the enablement channels, and required skillsets. In the absence of a sound strategy and an execution plan, the isolated analytics efforts can quickly go adrift since it would be almost impossible to ask the ``right'' questions. While this topic is not the focus here, it is still important to recognize the need for such strategy.
\item \emph{Culture change} to accept insights from \textit{validated}, \textit{verified} and \textit{principled} data-driven analytics into decision making at \textit{all} levels. Even though a lot of capabilities in analytics are being commoditized, it is extremely important that users, both the ones performing analytics and the ones ingesting the resulting insights, are aware of the assumptions and constraints of the methods applied, as well as the ranges in which these should be interpreted. Moreover, such a culture-change is not unidirectional. Data divisions also have the onus of understanding the domains and their operational constraints better to be able to deliver value and to complement the domain and engineering experts.
\item \emph{Innovation} to address open problems especially in the context of respective business applications. Organizations need to invest in innovation since differentiation will result from novel capabilities and well engineered integration.
\end{enumerate}

\section{Societal impact and areas of concerns}\label{sec:social}

While the technological feasibility of big data analytics for the IoT has been demonstrated in limited contexts, much more needs to be done to realize the broader vision. Not only the existing technology needs to be perfected, further innovation is needed to solve current bottlenecks as well as address longer term requirements. On the IoT end, this can mean increasing efficiency and affordability of data acquistion devices while reducing energy  consumption as well as standardization of M2M service layer. Efforts are also needed in building common communications standards (while efforts are underway we do not have any consensus yet) and improving interoperability across data, semantics and organizations. In addition to the sources listed earlier, see~\citep{vermesan13} for a discussion on some additional aspects of IoT as well as architectural approaches in different contexts.

On the big data processing and analytics, we have just scratched the surface. Improved solutions are needed for problems such as analyzing massive temporal data, automated feature discovery, robust learning, analyzing heterogeneous data, efficiently managing complex, as well as meta-data, performing real-time analytics and handling streaming data (see, for instance,~\citep{zhou14,zicari13}). 

However, from a social point of view, there are also some major areas of concern that need to be addressed. We broadly divide these concern areas into two categories. The ones in the first category are \textit{technological challenges}: research community has been sensitized to these and work is currently underway to better understanding and addressing them. However, it must be mentioned that these areas warrant more attention and effort than they currently receive. Main areas in this first category include:

\begin{enumerate}
\item \textit{Privacy Issues:} Machine learning and data mining communities as well as other fields including policy, security and governance have been working on these issues for some time. From an analytics perspective, privacy preserving data mining has developed into a subfield and considerable effort has gone into studying privacy challenges in data mining~\citep{matwin13,navarro14}, data publishing~\citep{Fung10} and, to some extent, integration and interactions of sensors~\citep{Aggarwal11}. However, these efforts have focused mainly on the data and analytics layers. Better protocols are also needed for other layers in the IoT stack. For instance, privacy and de-identification at the data acquisition layer needs to be efficiently addressed. For every application, there are also specific requirements, both regulatory and technological, that should govern privacy concerns. For instance, in the US, HIPAA\footnote{Health Insurance Portability and Accountability Act} governs the majority of the requirements in dealing with medical data in many scenarios. Clear data governance and handling policies are needed to guide the efforts in the desired direction.
\item \textit{Security Issues:} Security is always a concern in the case of large distributed systems. The more access points a network has, the more vulnerable it becomes. In the absence of clear and agreed upon standards and protocols, the security challenges are increased exponentially. In fact, security issues in the IoT are already a reality. For instance,~\citet{Witten14} discusses top security mishaps in various contexts of IoT. Some work in this direction is already being done (e.g.,~\citep{glas12,yavuz13}) and needs to continue and expand. 
\item \textit{Interpretability issues:} When employing analytics models in practice, we need to confirm how much we can rely on abstract models generalized based on non-linearities in the data and what aspects require interpretability of these models. Some requirements can be imposed due to the nature of application field (e.g., due to regulations) while in others interpretability can be needed to make use of the findings (e.g., gene identification). Sophisticated models can undoubtedly leverage more information from data compared to their simpler, interpretable counterparts. However, better evaluation and validation mechanism should be put in place to guarantee generalizability.
\item \textit{Data quality issues:} Often, it is seen that the acquired data does not support desired analysis. For example, in a lot of cases, the acquired data from sensors is not intended to performed inductive inference at scale but rather is aimed to target a specific aspect such as safety, or reliability. Such cases would need an enhanced understanding of what use can be made of available data in the analytics context and how data quality can be ascertained.
\end{enumerate}

The second category of issues is even more important in our opinion. We call these \textit{adoption challenges}, referring to the issues resulting from inevitable, pervasive and ubiquitous adoption of analytics in various domains. This should not be viewed as an argument against more integration of analytics. Just as any other technology, analytics is a neutral force and the implications of its integration and use would rely on responsible choices made while trying to leverage it. Our aim is to sensitize the community so as not to overlook these as we move towards a new paradigm. Even though it is not possible to have immediate answers, we would like to highlight the issues to raise awareness of them during decision making processes as well as evolving strategies:

\begin{enumerate}
\item \textit{Model reliability, validation and adaptation:} This is possibly the most widely discussed issues in the current list. Just by statistical chance, given that the models operate on vast amounts of data, correlations \emph{will} be found. How should these correlations be validated? Standardization and agreement is required to evaluate these models and understand the associated risks. Principled forward testing mechanisms will be required, especially in cases of rare events such as asset failures. Backward testing and validation set-based evaluations are limited. Further, robustness of the models needs to be ascertained in changing environments either via model adaptation or via regular evaluation and requirements caliberation. Moreover, as these models interact with the environment and do not operate in isolation, their validation and verification becomes all the more crucial. This is especially important since the cost of doing ``wrong'' analytics may be significant for certain areas such as physical and mission-critical systems. 
\begin{enumerate}
\item \textit{The risk of over-sophistication:} Extreme fine-tuning may result in models that can be very effective, but only for a very short period. If analytics has to be integreted into the process, it should be long lasting and adaptable. This requires more than just models that take into account evolution of the data or labels (e.g. concept drift) but also refers to how these models are utilized, how the expectations change over time and how the process responds to the results.
\end{enumerate}
\item \textit{Integration and reconciliation with our physical understanding of the world:} As IoT grows, analytics will increasingly be integrated in the environment, whether embedded in devices or assisting in decision making based on aggregate analytics. It is extremely important that we can reconcile these capabilities with the basis that we use to build and operate the physical devices (e.g., physics-based models). 
An argument can be made to restrict the models to ``interpretable'' ones when it comes to analytics. However, this trades off the knowledge that can be had from non-linear models in deriving non-obvious relationships hidden in the data. We need better mechanisms to integrate these models and to validate their findings.
\item \textit{Human-analytics interaction:} As technology becomes pervasive, it tends to have an \textit{assumed truth effect}, meaning that over time the users take the results with ever increasing trust. Consequently, in scenarios where decision-making will move closer to automated approaches, we should be mindful of their advantages as well as limitations. For instance, automated approaches have the potential to reduce the variations resulting from manual approaches. However, in some cases such variations are desired, even required, so that we can advance our understanding through a multitude of perspectives. It is timely to start seriously discussing about how humans will interact with analytics moving forward; how would this impact the decision making; would this lead to undesired uniformity?; will we be able to notice inconsistencies and errors in the suggested decisions as our reliance on these models increase? How would these models respond to evolving realities of the world? How would the automated decision making impact policy? 
\item \textit{Potential for systemic errors and failures:} Another aspect to consider is how much of a threat do automated decision making models pose to systematic as well as systemic failures as they become pervasive; Can the errors of individual pieces multiply resulting in system-wide risks? Will they have potential to bring down the whole system? Can massive interconnectivity result in a system-wide spread of failures, threats or even attacks? Note that the individual risks can be small and gradual but taken together they may lead to serious implications. Consequently, a risk containment mechanism will be needed in interconnected systems.
\begin{enumerate}
\item \textit{Localization of ``failures'':} If system-wide events were to happen, would we have the ability to locate the sources? will we be able to quarantine a part of the system? Moreover, what effect would this have on the users since these systems will be an integrated part of peoples' lives? How would the necessary and important services be affected?
\end{enumerate}
\item \textit{Personalization vs. Limitation of choice:} There can be intended and unintended, but nonetheless undesirable, consequences of ``personalization'' of services to individual lifestyles. On the unintended side, can over-personalization limit choice? For instance, as an effort to recommend the most relevant options, a subset of possible options is presented to the user. However, over time, and with increasing reliance on these recommendations, the users' exposure to possibilities outside of these recommendation-ranges can potentially be adversely impacted.  Such systems can then potentially be used for malicious purposes such as social engineering around issues. Just as policy should take into account these aspects as technology grows, technologists also share the responsibility to contribute to addressing these issues.
\end{enumerate}

\section{Concluding Remarks}\label{sec:conclusion}

In this chapter, we discussed how big data technologies and the internet of things are playing a transformative role in the society. The pervasive and ubiquitous nature of such technologies will profoundly change the world as we know it, just as the industrial revolution and the internet did in the past. We discussed opportunities in various domains both from an industrial and from the consumers' perspective. Given the data acquisition capabilities that are in place in the context of monitoring physical assets, the immediate opportunities are bigger from an industrial perspective. On the consumer end, we are currently undergoing a transformation as physical devices capable of advanced sensing become part of our routine life. Consumer applications will start witnessing a rapid growth in integrated services and systems, which we believe will generate much more value in contrast to one-off offerings as noted in Section~\ref{subsec:integratedsys}, once a critical mass of such interconnected devices is reached in various domains. The capabilities in leveraging big data in both of these contexts are already transitioning from performing \textit{descriptive analytics} to \textit{predictive analytics}. For instance, based on real-time sensor data, we can predict certain classes of field events (e.g., failures or malfunctions) for heavy assets such as aircraft engines and turbines more reliably; this complements physics-based models employed in such cases. As these technologies mature, they will enable another transition from predictive to \textit{prescriptive analytics} whereby recommendations on resolutions of such events could be made. This may develop to the extent of devices themselves taking corrective actions, and thus making them self-aware and self-maintaining. Even though we are already witnessing a paradigm shift, more needs to be done on various fronts, such as advancements in big data technologies, analytics, privacy and security, and policy making. In addition, the requirements at an organizational level in terms of readiness to harness the value resulting from analytics are discussed. 

We then discussed broad social implications and highlighted areas of concerns as these technologies become pervasive. We organized these concerns into two categories: \textit{technological challenges} that are relatively better understood, even if not entirely resolved, and \textit{adoption challenges} that we believe are more unclear. As the adoption and integration of such technologies grows, so will our understanding of the implications evolve. However, the pace of change is fast indeed, and we will need to be quick in understanding this evolving landscape, analyzing the resulting changes and defining proper policies and protocols at various levels. Factors such as \textit{human-analytics interaction} will also play an important role in how responsibly and effectively analytics complement our decision-making ability as well as how much autonomy these systems eventually obtain.

Finally, we should reiterate that technologies are neutral. Any technology will have implications on society. The onus is on us to define how the technology is adopted in a responsible manner.

\section*{Appendix}

Links to entities referred to in the article (in alphabetical order):
\begin{itemize}
\item Amazon AWS for IoT: \url{http://aws.amazon.com/iot/}
\item Amazon Echo: \url{http://www.amazon.com/oc/echo/ref_=ods_dp_ae}
\item Beddit: \url{http://www.beddit.com/}
\item Being: \url{http://www.zensorium.com/being} 
\item Bosch, ABB, LG and Cisco's joint venture announced recently to cooperate on open standards for smart homes: \url{http://www04.abb.com/global/seitp/seitp202.nsf/0/9421f99d7575ceccc1257c1d0033fa4a/\$file/8364IR_en_Red_Elephant_20131024_final.pdf}
\item Bosch Indego: \url{https://www.bosch-indego.com/gb/en/}
\item Cloud Foundry: \url{http://www.cloudfoundry.org/about/index.html}
\item Cloudera and Hortonwork's real-time offering: \url{http://www.infoq.com/news/2014/01/Spark-Storm-Real-Time-Analytics}
\item Ego LS: \url{http://www.liquidimageco.com/}
\item Fitbit: \url{http://www.fitbit.com/}
\item Hadoop Ecosystem: See, for instance,~\url{http://hadoopecosystemtable.github.io/}
\item Hubject, a joint networking mobility initiative of the BMW group, Bosch, Daimler, EnBW, RWE and Siemens: \url{https://www.bosch-si.com/solutions/mobility/our-solutions/hubject.html}
\item IBM SyNAPSE: \url{http://www.research.ibm.com/cognitive-computing/neurosynaptic-chips.shtml#fbid=CAQQuy4xAkK}
\item Jawbone: \url{https://jawbone.com/}
\item Jibo: \url{http://www.myjibo.com/}
\item Microsoft's IoT offerings: \url{http://www.microsoft.com/en-us/server-cloud/internet-of-things.aspx}
\item NVidia's Tegra X1: \url{http://www.nvidia.com/object/tegra-x1-processor.html}
\item Pacif-i: \url{http://bluemaestro.com/}
\item Pandas: \url{http://pandas.pydata.org/}
\item Predictive Model Markup Language (PMML): \url{http://www.dmg.org/v4-1/GeneralStructure.html}
\item Qualcomm Zeroth: \url{https://www.qualcomm.com/news/onq/2013/10/10/introducing-qualcomm-zeroth-processors-brain-inspired-computing}
\item Spark: \url{https://spark.apache.org/}
\item Sproutling: \url{http://www.sproutling.com/}. 
\item Storm:\url{https://storm.apache.org/}
\item Tado: \url{https://www.tado.com/}
\item Tzoa: \url{http://www.mytzoa.com/#homepage}
\item Vessyl: \url{https://www.myvessyl.com/}
\item Withings Aura: \url{http://www.withings.com/us/withings-aura.html}
\item Zementis: \url{http://zementis.com/}
\end{itemize}



\bibliography{mohak-bib}
\bibliographystyle{plainnat}
\end{document}